\documentclass[sigconf]{acmart}

\AtBeginDocument{%
  \providecommand\BibTeX{{%
    \normalfont B\kern-0.5em{\scshape i\kern-0.25em b}\kern-0.8em\TeX}}}

\setcopyright{acmcopyright}
\copyrightyear{2023}
\acmYear{2023}
\acmDOI{XXXXXXX.XXXXXXX}

\acmConference[Conference acronym 'XX]{Make sure to enter the correct
  conference title from your rights confirmation emai}{June 03--05,
  2018}{Woodstock, NY}
\acmPrice{15.00}
\acmISBN{978-1-4503-XXXX-X/18/06}

\usepackage{multirow}
\usepackage{svg}
\usepackage{subcaption}
\usepackage{threeparttable}
\usepackage[normalem]{ulem}
\usepackage{enumitem}
\useunder{\uline}{\ul}{}
\newtheorem{mydef}{Definition}
\AtBeginDocument{%
  \providecommand\BibTeX{{%
    \normalfont B\kern-0.5em{\scshape i\kern-0.25em b}\kern-0.8em\TeX}}}

\copyrightyear{2023}
\acmYear{2023}
\setcopyright{acmlicensed}\acmConference[CIKM '23]{Proceedings of the 32nd ACM International Conference on Information and Knowledge Management}{October 21--25, 2023}{Birmingham, United Kingdom}
\acmBooktitle{Proceedings of the 32nd ACM International Conference on Information and Knowledge Management (CIKM '23), October 21--25, 2023, Birmingham, United Kingdom}
\acmPrice{15.00}
\acmDOI{10.1145/3583780.3614902}
\acmISBN{979-8-4007-0124-5/23/10}

\begin{document}

\title{Group Identification via Transitional Hypergraph Convolution with Cross-view Self-supervised Learning}

\author{Mingdai~Yang}
\affiliation{%
  \institution{University of Illinois at Chicago}
  \city{Chicago}
  \country{USA}}
\email{myang72@uic.edu}

\author{Zhiwei~Liu}
\affiliation{%
  \institution{Salesforce AI Research}
  \city{Palo Alto}
  \country{USA}
}
\email{zhiweiliu@salesforce.com}


\author{Liangwei~Yang}
\email{lyang84@uic.edu}
\affiliation{%
  \institution{University of Illinois at Chicago}
  \city{Chicago}
  \country{USA}}

\author{Xiaolong~Liu}
\email{xliu262@uic.edu}
\author{Chen~Wang}
\email{cwang266@uic.edu}
\affiliation{%
  \institution{University of Illinois at Chicago}
  \city{Chicago}
  \country{USA}}

\author{Hao Peng}
\affiliation{%
   \institution{School of Cyber Science and Technology, Beihang University,}
   \country{Beijing, China}\\
   \institution{Yunnan Key Laboratory of Artificial Intelligence, Kunming University of Science and Technology,}
   \country{Kunming, China}}
\email{penghao@buaa.edu.cn}
\authornote{Corresponding author}

\author{Philip S.~Yu}
\affiliation{%
  \institution{University of Illinois at Chicago}
  \city{Chicago}
  \country{USA}}
\email{psyu@uic.edu}

\renewcommand{\shortauthors}{Mingdai Yang et al.}

\begin{abstract}
 With the proliferation of social media, a growing number of users search for and join group activities in their daily life. This develops a need for the study on the group identification (GI) task, i.e., recommending groups to users. 
 The major challenge in this task is how to predict users'  preferences for groups based on not only previous group participation of users but also users' interests in items. 
 Although recent developments in Graph Neural Networks (GNNs) accomplish embedding multiple types of objects in graph-based recommender systems, they, however, fail to address this GI problem comprehensively.
In this paper, we propose a novel framework named \textbf{G}roup Identification via \textbf{T}ransitional Hypergraph Convolution
with \textbf{G}raph \textbf{S}elf-supervised Learning (GTGS). 
We devise a novel transitional hypergraph convolution layer to leverage users’ preferences for
items as prior knowledge when seeking their group preferences. To construct comprehensive user/group representations for GI task, we design the cross-view self-supervised learning to encourage the intrinsic consistency between item and group preferences for each user, and the group-based regularization to enhance the distinction among group embeddings.
Experimental results on three benchmark datasets verify the superiority of GTGS. 
Additional detailed investigations are conducted to demonstrate the effectiveness of the proposed framework.
 
\end{abstract}



\maketitle

\section{Introduction}

With the development of online shopping, information collection and decision-making have become essential yet overwhelming for individual customers. 
To seek suggestions for reference before purchasing, a growing number of users decide to join groups on online platforms and communities.
For example, an expectant mother would seek suggestions from other experienced mothers in online communities. 
Online groups provide a space for users to discuss their demands and share their experiences. 
Moreover, online groups can affect and develop users' interests. 
For example, a group of movie fans would discuss their anticipated movies, and a group of video game players would debate whether a newly released game is worth buying. 
In terms of platforms, users' participation and retention rate can be boosted by the attachment to online groups~\cite{engagement@17}.
Therefore, it improves the long-term stickiness of users to platforms if personalized social groups can be suggested for users.
In this paper, we define this problem as \textit{group identification}\footnote{We use \textit{group identification} rather than \textit{group recommendation} is because the later usually refers to recommending items to a group of users.}~(GI) task, which aims at recommending groups to users. 


If regarding users and groups as nodes and their interactions as edges, we may thus resolve the group identification~(GI) task as predicting edges over the user-group bipartite graph~\cite{ZhaHDGS01,LiC13}.
Then, existing graph-based recommender systems methods~\cite{lightgcn20,sgl,enmf20,yang2021consisrec} would be adopted, which predict edges based on interactions in the graph.
However, GI task is far more complex than this because the group participation of users also depends on the item interests of users. 
For instance, a user may join a group because other members purchased the item of her interest. 
Therefore, simultaneously capturing collaborative signals among users, groups, and items is necessary for accurate recommendation in GI task.


The recent developments of graph neural networks (GNNs) have achieved great success in embedding multiple types of nodes in graph-based recommender systems~\cite{hgcncc,dhcf,mhcn,yang2023ranking,yang2022large}.
To name a few,
HetGNN~\cite{hetergnn1} integrates Bi-LSTM~\cite{HochreiterS97} to capture both structure and content heterogeneity in graphs.
GTN~\cite{hetergnn2} learns from a soft selection of edge types and composite relations in heterogeneous graphs through generative meta paths.
HGT~\cite{hetergnn3} proposes heterogeneous mutual attention to characterize the attention over different types of edges for modeling heterogeneity.
However, edges in most existing graphs only characterize the pair-wise semantics, which contradicts the fact that users join a group as a unified whole. 
To be more specific, a group is a union of all users in this group. A new user joins this group because of his or her potential relationships with all users in the groups. 
In this sense, pair-wise connections cannot explicitly express the \textit{union semantics} in groups. 

\begin{figure}
  \centering
  \includegraphics[width=0.85\linewidth]{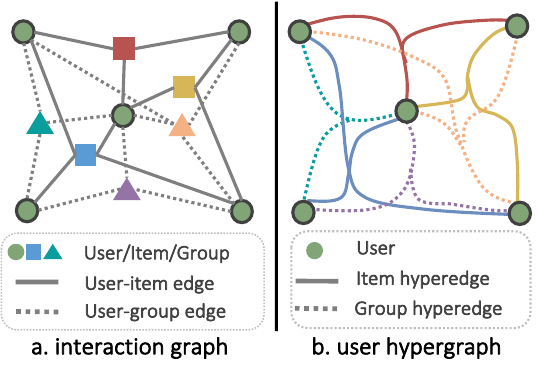}
  \caption{Toy example of how to construct a GI hypergraph from user-group-item interaction graph. }
  \label{illustration}
\end{figure}

To this end, we adopt hypergraphs~\cite{ZhouHS06,hnn} for GI task, where hyperedges connect multiple nodes synchronously.
We illustrate a toy example of the hypergraph construction for GI task in Figure~\ref{illustration}.
We construct a GI hypergraph from a user-item-group interaction graph. 
The hyperedges are items and groups, which are represented as solid and dash lines in Figure~\ref{illustration}(b), respectively. 
The hypergraph structure intrinsically reflects the union semantics of users.
For example, the bottom purple triangular in the interaction graph denotes a group, which connects to three users through three edges, whereas, in the GI hypergraph, we convert this group to a hyperedge, denoted as purple dashline, which synchronously connects three nodes, and thus representing the union of three users.  

However, previous hypergraph convolution is defective in propagating information on hypergraphs, although it has been widely applied to aggregate information in recommender systems~\cite{dhcf,hgcncc}.
In previous methods, the information from nodes in a hypergraph is aggregated to hyperedges and then directly aggregated back to those nodes by hypergraph convolution. 
They neglect the intrinsic hyperedge information during aggregation, thus being suboptimal for information propagation.
For example, if we construct hyperedges to connect the groups joined by the same users in GI problem, previous hypergraph convolution only builds the representations of those groups based on their member composition, but completely ignores the item preferences of their members. 
Nonetheless, in most e-commercial platforms, the content of a group is up to the popular items attracting the members. 
For item-related group identification without group-item interaction data available, it is challenging to transmit information on members' item preferences to groups by hypergraph learning.


The other challenge in applying hypergraph learning to GI task is how to harmonize the information learned from different hyperedges for the same nodes, i.e., the information from different views of hypergraphs. For example, although a user's item preferences and group preferences can be modeled through item and group hyperedges discriminatively in the user hypergraph, without integrating these two types of preferences coherently, the recommender system is still unable to deduce her group preferences based on her item preference.
We argue that it is crucial to maintain the intrinsic consistency between users' item and group preferences during hypergraph learning. Given no available group-item interaction, this consistency inside each user can be regarded as an alternative group-item connection, which provides necessary information for recommending groups to users. 

To this end, we propose a novel framework called Group Identification via Transitional Hypergraph Convolution
with Graph Self-supervised Learning (GTGS). GTGS is a hypergraph neural network built upon one user-view group hypergraph, and two item-view and group-view user hypergraphs.
We propose a novel transitional hypergraph convolution (THC) layer to generalize existing hypergraph convolution. 
THC layer endows the model with the ability of characterizing the both hyperedge information propagated from nodes and intrinsically existing in hyperedges.
In this way, the intrinsic hyperedge information on users' preferences for items is transmitted to groups. 
Regarding learning user representation from both group-views, we leverage contrastive self-supervised learning (SSL) to encourage the consistency between item and group preferences of users. 
After convolution on item-view and group-view user hypergraphs, we obtain two views of embeddings for all users. 
Then, we develop a novel cross-view self-supervised learning (CSSL) based on InfoNCE contrastive loss~\cite{infonce} to maximize the mutual information between two views of the same user in a self-supervised manner.
Additionally, we propose a new group-based regularization term to alleviate the embedding collapse issue for contrastive learning. 
This group-based regularization is leveraged to enhance the distinction between different groups for constructing distinguishable group representations. The main contributions of this paper are summarized as follows:
\begin{itemize}[leftmargin=*]
    \item  We propose a novel framework GTGS, a model based on hypergraph learning for group identification, which leverages hyperedges to express the union semantics in users and groups.
    \item We devise a novel THC layer to harness the intrinsic information on hyperedges, which generalizes existing hypergraph convolution and GCN-based models. 
    \item We design a novel CSSL paradigm to optimize user and group representations respectively and propose a group-based regularization to improve the distinction of group embeddings.
    \item We conduct extensive experiments on three real-world datasets. The significant improvement of GTGS on all datasets indicates its superiority in tackling the GI problem.
\end{itemize}


\section{Related works}
\subsection{Group Recommender Systems}
Group recommender systems refer to recommending groups to their potential members. 
Traditional group recommender systems apply various algorithms to recover user-group membership matrices with available side information.
For example, semantic information from descriptions of groups~\cite{chen08} and visual information from photos shared by users~\cite{wang12} can be incorporated with a collaborative filtering framework to perform personalized group recommendations.
User behaviors in different time periods~\cite{wang16, yang21}, such as joining groups, can also be leveraged for recommending groups to users. 
However, the requirement for side information degrades the performance of those methods when recommending groups to users with only interaction information. 
Some recent works~\cite{grouprec21, Telegram21} investigate recommending groups to users with only user-group interactions
, while item interaction information is ignored.

Besides recommending groups to users, the term \textit{group recommendation} in literature also refers to recommending items to a group of users~\cite{groupim20, agree18}, which differs from the focus in this paper.

\subsection{Hypergraph Learning Based Recommendation}
Inspired by the success of hypergraph learning in other graph-based tasks~\cite{feng19,gao22}, the hypergraph is also introduced in recommender systems due to its capability of modeling complex high-order dependencies through hyperedge-node connections. HGCN-CC~\cite{hgcncc} improves HGCN~\cite{feng19} with feature crossing and contrastive learning for the user-item recommendation. DHCF~\cite{dhcf} utilizes residual connections to consider original features and aggregated related representations simultaneously during modeling the hybrid multi-order correlations between users and items.
Hypergraphs can also be constructed based on side information in recommender systems. As an early attempt, MRH~\cite{mrh} builds hyperedges among users, music, and the music tag to construct a hypergraph for music recommendation. MHCN~\cite{mhcn} proposes
a multi-channel hypergraph convolutional network to enhance
social recommendation by leveraging high-order user-user relations.
HMF~\cite{hmf} leverages hypergraph structure to describe the interior relation of the social network.

To the best of our knowledge, no previous work has applied hypergraph learning for group recommender systems. Since the hyperedge simultaneously connecting multiple nodes is a natural structure for handling groups of users in recommender systems, some recent works leverage hypergraph learning to predict potential items for given groups of users. HHGR~\cite{hhgr} designs hierarchical hypergraph convolution on fine- and coarse-grained hypergraphs. HCR~\cite{hcr} defines an overlap graph of a hypergraph to learn the group’s general preference independent of its members' interests.

\subsection{Recommendation with Self-Supervised Learning}
Self-supervised learning is an emerging paradigm to generate additional supervised signals for improving representation quality. It has reached most recommendation topics and achieved remarkable success. Recommendation models with self-supervised learning can be generally categorized into two branches according to the type of self-supervised learning task: generative methods and contrastive methods. Inspired by the masked language models~\cite{bert}, generative methods learn by reconstructing the input data from its corrupted version~\cite{bert4rec,gbert}.
Contrastive methods compare multiple views of input data and contrast positive samples with negative samples to learn discriminative representations~\cite{sgl,dhcn}. 

The construction of high-quality self-supervised signals plays a pivotal role in self-supervised learning~\cite{TianSPKSI20}. Most previous works~\cite{bert4rec,sgl,CFM} augment the data to create multiple views as self-supervised signals. 
However, complex data augmentation cost additional space and time Recent works~\cite{Lee0P22,simgcl} verify that common data augmentations can bring negative impacts for graph-based tasks. In this work, we design two novel augmentation-free self-supervised learning paradigms to optimize user and group representations.

\section{Preliminaries}
In this section, we first formulate the problem of group identification (GI), then introduce three hypergraphs for GI and their construction.  
\begin{mydef}
\textbf{(Ranking-based Group Identification)}. Given three disjoint node sets, including a user set $\mathcal{U}$, a group set $\mathcal{G}$ and an item set $\mathcal{I}$, and the interactive edges, i.e., user-group edges $E_{\mathcal{U},\mathcal{G}}$ and user-item edges $E_{\mathcal{U},\mathcal{I}}$, an interaction graph is defined as $\mathcal{T}=(V,E)$, where $V = \mathcal{U} \cup \mathcal{I} \cup \mathcal{G}$ and $E = E_{\mathcal{U},\mathcal{G}} \cup E_{\mathcal{U},\mathcal{I}}$.
The group identification (GI) for a user $u$ is to predict a ranking list of groups $\{g_1, g_2, \dots, g_k\}$, with which this user has no interactions in the graph $\mathcal{T}$.
\end{mydef}
In other words, we recommend a list of groups that this user $u$ is of potential interest in GI. Note that we distinguish the group as another entity rather than a simple union of users due to its special characteristics, e.g., group information.

\begin{mydef}
\textbf{(User-view Group Hypergraph).} The user-view group hypergraph of an interaction graph $\mathcal{T}$ is denoted as $\mathcal{T}^{u}_{g}=(\mathcal{G}, \mathcal{E}_{u})$, where groups $\mathcal{G}$ are nodes and $\mathcal{E}_{u}$ are hyperedges. 
We let users be hyperedges. 
An incidence matrix $\mathbf{H}\in\{0,1\}^{|\mathcal{G}|\times |\mathcal{U}|}$ is used to represent connections among group nodes.
\end{mydef}

Analogously, we define the \textbf{Group-view User Hypergraph} and \textbf{Item-view User Hypergraph}, denoted as $\mathcal{T}_{u}^g$ and $\mathcal{T}_{u}^i$, respectively, whose nodes are users and hyperedges are groups and items, respectively.
$\mathcal{T}_{u}^g$ and $\mathcal{T}_{u}^i$ are associated with two incidence matrices, denoted as $\mathbf{U}_g\in\{0,1\}^{|\mathcal{U}|\times |\mathcal{G}|}$ and $\mathbf{U}_i\in\{0,1\}^{|\mathcal{U}|\times |\mathcal{I}|}$, respectively. 
Figure~\ref{fig:framework} presents an example of constructing the above hypergraphs from an interaction graph.

\section{Methods}
\begin{figure*}[h]
  \centering
    \includegraphics[width=\linewidth]{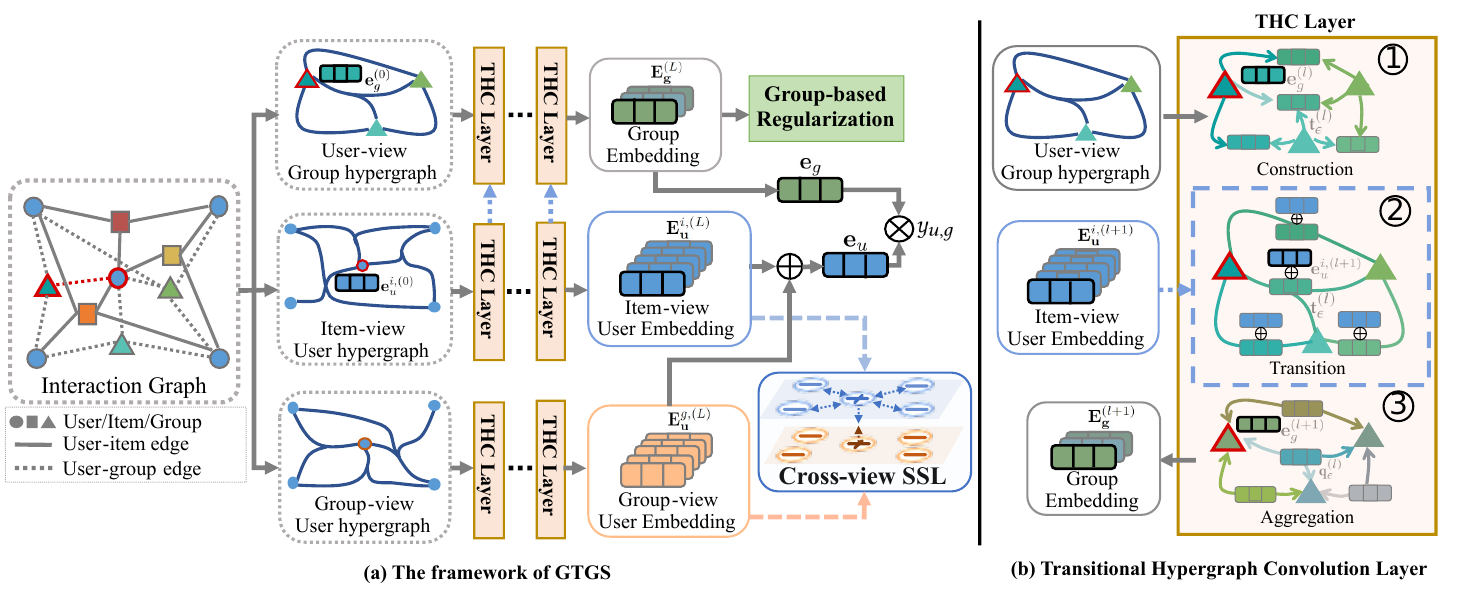}
  \caption{(a) The overall framework of GTGS. 
  First, we construct three hypergraphs and apply THC layers to them. Next, we conduct group-based regularization on the output group embeddings, and employ cross-view SSL to optimize item-view and group-view user embeddings. At last, the inner product of group embedding and item-view user embedding is calculated for prediction; 
  (b) the illustration of the Transitional Hypergraph Convolution (THC) layer.}
  \label{fig:framework}
\end{figure*}

In this section, we present the proposed GTGS model for the group identification task. The framework of GTGS is shown in Figure~\ref{fig:framework}. We start by introducing all embedding layers to be trained in this framework.
Thereafter, we demonstrate how to leverage users' preferences for items as prior knowledge in group representation learning. 
Specifically, we adopt a THC layer to integrate users' preferences learned in the item-view user hypergraph into hyperedge features, and then aggregate these hyperedge features to group nodes. 
We further design SSL tasks to optimize 
 user and group representations respectively. We propose CSSL to encourage the intrinsic consistency between item-view and group-view user embeddings, and leverage group-based regularization to enhance the distinction between different group embeddings. Lastly, we conduct theoretical analyses on THC and prove it as a general case of graph convolution and hypergraph convolution.

\subsection{Embedding layer}
We maintain an embedding layer $\mathbf{E}\in \mathbb{R}^{d\times(2|{\mathcal{U}}|+|\mathcal{G}|)}$, where $d$ is the feature dimension and columns represent all trainable embeddings for users and groups in the three hypergraphs. Each user has two embeddings corresponding to the item-view and group-view user hypergraphs. We denote the initial item-view user embedding as $\mathbf{E}_\mathbf{u}^{i,(0)}$, the initial group-view user embedding as $\mathbf{E}_\mathbf{u}^{g,(0)}$, and the initial group embedding as $\mathbf{E}_\mathbf{g}^{(0)}$.

\subsection{Transitional Hypergraph Convolution}
Hypergraph convolution is widely used for information propagation among nodes through
hyperedges in hypergraphs. 
However, existing hypergraph convolution methods directly aggregate the information from neighbors connected by hyperedges, while neglecting intrinsic information of hyperedges. 
Hence, we devise a novel Transitional Hypergraph Convolution (THC) layer to resolve this issue. 
Specifically, before information propagated from node to node, we first construct the hyperedge information from connected nodes. We denote this initial step as information \textit{construction} of THC layer.  
Then, we propose an additional \textit{transition} step, which enables the fusion of hyperedge information from other resources.
Finally, in the final \textit{propagation} step, fused information is aggregated to nodes for inference of node embedding.
Next, to be more concrete, we present the calculation details of a THC layer over a user-view group hypergraph. The illustration is in Figure~\ref{fig:framework}(b).


\subsubsection{Information Construction} In the user-view group hypergraph, group nodes are connected by user hyperedges.
Therefore, we construct the information of a hyperedge by aggregating embeddings of  all nodes connected by this hyperedge, which is illustrated as in \textcircled{1} of Figure~\ref{fig:framework}(b). 
The information construction of a hyperedge $\epsilon$ is formulated as follows:
\begin{equation}\label{eq1}
    \mathbf{t}_\epsilon^{(l)} = \frac{1}{|\mathcal{N}_\epsilon|}\sum_{g'\in\mathcal{N}_\epsilon}\mathbf{e}_{g'}^{(l)},
\end{equation}
where $\mathbf{t}_\epsilon^{(l)}$ denotes temporary embedding of the hyperedge $\epsilon$, $\mathcal{N}_\epsilon$ is the set of group nodes connected by this hyperedge, and $\mathbf{e}_{g'}^{(l)}$ is the embedding of those group nodes as input of the $l$-th THC layer.
Eq.~(\ref{eq1}) represents constructing the hyperedge information via mean-pooling of its connected nodes.
We leave other variants such as max-pooling to future work.  

\subsubsection{Instrinsic Information Transition} In this step, we fuse the constructed information with intrinsic information of a hyperedge. 
Let the intrinsic embedding of a hyperedge  $\epsilon$ be $\mathbf{c}_{\epsilon}$, we fuse $\mathbf{c}_{\epsilon}$ with $\mathbf{t}_\epsilon^{(l)}$ via a \textit{transition layer} 
as follows:
\begin{equation}\label{eq:transition layer}
    \mathbf{q}_\epsilon^{(l)} = \text{Transition}(\mathbf{t}_\epsilon^{(l)}, \gamma\mathbf{c}_{\epsilon}^{(l)}),
\end{equation}
where $\mathbf{q}_\epsilon^{(l)}$ is the fused hyperedge embedding, and $\gamma$ is a scalar hyper-parameter to control the transition intensity. 
For example, a hyperedge in user-view group hypergraph is a user. 
Hence, the intrinsic information of the hyperedge is from the user embedding.
In this paper, we use add operation as the transition layer and leave the investigation of other variants to future work. 
In \textcircled{2} of Figure~\ref{fig:framework}(b), we demonstrate this step via adopting the intrinsic information of user hyperedges from the item-view user embeddings, \textit{i.e.,} $\mathbf{C}^{(l)} = \mathbf{E}_\mathbf{u}^{i,(l+1)}$, which will be introduced later. 


\subsubsection{Final Aggregation} We learn the group embedding by aggregating the fused embedding from all its connected hyperedges.
For a target group $g$, the aggregation step is formulated as follows:
\begin{equation}\label{eq3}
    \mathbf{e}_{g}^{(l+1)} = \frac{1}{|\mathcal{N}_g|}\sum_{\epsilon\in\mathcal{N}_g}\mathbf{q}^{(l)}_{\epsilon},
\end{equation}
where $\mathbf{e}_{g}^{(l+1)}$ denotes the output group embedding, $\mathcal{N}_g$ is the set of hyperedges connected to this group, and $\mathbf{q}^{(l)}_{\epsilon}$ is the fused embedding of its connected hyperedges from Eq.~(\ref{eq:transition layer}).
Again, we only investigate the mean-pooling aggregation of hyperedge information and leave other variants to future study. 
This step is demontrated in Figure~\ref{fig:framework}(b) \textcircled{3}. 


\subsubsection{THC in Matrix Form}. 
To offer a holistic view
of convolution, we formulate the matrix form of transitional hypergraph convolution (equivalent to Eq.~(\ref{eq1})-(\ref{eq3})) as:
\begin{equation}\label{thc}
\begin{aligned}
    \mathbf{E}_{\mathbf{g}}^{(l+1)} &= \text{THC}(\mathbf{E}_\mathbf{g}^{(l)},\mathbf{H},\gamma\mathbf{C}_\epsilon^{(l)})\\
    &= \mathbf{D}^{-1}\mathbf{H}\cdot
    \text{Transition}(\mathbf{B}^{-1}\mathbf{H}^\top\mathbf{E}_{\mathbf{g}}^{(l)}, \gamma\mathbf{C}^{(l)}),
\end{aligned}
\end{equation}
where  $\mathbf{E}_{\mathbf{g}}^{(l)}$ is the input embedding from $l$-th layer, $\mathbf{H}$ is the incidence matrix, $\mathbf{C}^{(l)}$ denotes intrinsic hyperedge information, and  $\mathbf{E}_{\mathbf{g}}^{(l+1)}$ is the output. $\mathbf{D}$ is the degree matrix of nodes and $\mathbf{B}$ is degree matrix of hyperedges for normalization. 
THC layer is a more general version of hypergraph convolution, which degrades to existing hypergraph convolution~\cite{mhcn} if let $\gamma = 0$ and dismiss the Transition layer. 

\subsection{Cross-view Self-supervised Learning}
Recall that we construct two user hypergraphs, \textit{i.e.,} \text{item-view} and \text{group-view} user hypergraph. 
In this section, we introduce how to infer user embeddings from both hypergraphs and how to harness them via contrastive self-supervised learning simultaneously. 
\subsubsection{User Embeddings}
We employ $L$ THC layers on both \text{item-view} and \text{group-view} user hypergraphs to infer the item-view user embeddings $\mathbf{E}_{u}^{i,(L)}$ and the group-view user embeddings $\mathbf{E}_\mathbf{u}^{g,(L)}$ as:
\begin{equation}\label{eq:information_propagation}
\begin{aligned}
\mathbf{E}_\mathbf{u}^{i,(L)} &= \text{THC}^{L}(\mathbf{E}_\mathbf{u}^{i,(0)},\mathbf{U}_i,0), \\
\mathbf{E}_\mathbf{u}^{g,(L)} &= \text{THC}^{L}(\mathbf{E}_\mathbf{u}^{g,(0)},\mathbf{U}_g,0),
\end{aligned}
\end{equation}
where $\mathbf{U}_i$ and $\mathbf{U}_g$ are incidence matrices. $\mathbf{E}_\mathbf{u}^{i,(0)}$ and $\mathbf{E}_\mathbf{u}^{g,(0)}$ are the input user embeddings.
Note that here we eliminate the intrinsic hyperedge information for both views by setting the $\gamma = 0$ in Eq.(\ref{eq:transition layer}), which retrains simplicity of the model.
Intuitively, after $L$ THC layer, the item-view and group-view user embedding characterize the collaborative signals in user-item interaction and user-group interactions, respectively. 
In the following paper, we ignore the superscript $(L)$ for legibility and simply use $\mathbf{E}_\mathbf{u}^{i}$ and $\mathbf{E}_\mathbf{u}^{g}$ to denote the final embedding from item-view and group-view, respectively. 
We calculate the final user embeddings via a weighted sum the embeddings from two views as:
\begin{equation}\label{eq:user embed}
    \mathbf{E}_{u} = \beta \mathbf{E}_\mathbf{u}^{i} + (1-\beta)\mathbf{E}_\mathbf{u}^{g},
\end{equation}
where $\beta$ is a hyper-parameter to balance two views. 
\subsubsection{Cross-view Self-supervised Learning}
Though we represent the final user embedding as the sum of two views, the consistency between the two views is however neglected. 
The intuition is that we should encourage the model to recommend similar groups to users who prefer similar items, and vice versa. 
We devise a Cross-view Self-Supervised Learning (CSSL) task to enhance the consistency between item-view and group-view user representations as shown in Figure~\ref{fig:framework}(a). 
To establish consistency between the user representations under the two views, we deploy the InfoNCE loss to maximize the mutual information between the two views:
\begin{equation}\label{eq:cssl}
    \mathcal{L}_{cssl}^{user} = \sum_{u=0}^{|\mathcal{U}|}-\log\frac{\text{exp}(\text{sim}(\mathbf{e}_{u}^{i},\mathbf{e}_{u}^{g})/\tau_\mathbf{u})}
    {\sum_{v=0}^{|\mathcal{U}|}\text{exp}(\text{sim}(\mathbf{e}_{u}^{i},\mathbf{e}_{v}^{g}))/\tau_\mathbf{u})},
\end{equation}
where sim($\cdot, \cdot$) measures the cosine similarity between two vectors and $\tau_u$ is a temperature hyper-parameter. 
The item-view embedding $\mathbf{e}_{u}^{i}$ and group-view embedding $\mathbf{e}_{u}^{g}$ is interpreted as users' preference towards items and groups.

This loss function encourages the agreement between the two views of the same user and the distinction between the two views of different users. 
In other words, the consistency of the two views is maximized for each individual user.

\subsubsection{Group-based Regularization} 
Inspired by recent works~\cite{jing2021understanding,qiu2022contrastive} investigating the embedding collapse problem in contrastive learning, we  
propose a group-based regularization to improve the distinction of group representations. 
The reason of the embedding collapse issue in our loss function Eq.~(\ref{eq:cssl}) is that the excessive focus on the closeness of interacted user-group pairs results in a trivial solution of group embeddings collapsing to users who joined plenty of groups.
Thus, we apply a regularization to encourage all group embeddings to be uniformly distributed in the space. 
Concretely, we propose a new variant of the InfoNCE loss to regularize the distribution of group embeddings:
\begin{equation}
    \mathcal{L}_{reg}^{group} = \sum_{g=0}^{|\mathcal{G}|}-\log\frac{\text{exp}(1/\tau_\mathbf{g})}
    {\sum_{k=0}^{|\mathcal{G}|}\text{exp}(\text{sim}(\mathbf{e}_g,\mathbf{e}_k)/\tau_\mathbf{g})},
\end{equation}
where $\mathbf{e}_g$ and $\mathbf{e}_k$ are group embeddings learned by the THC layer, and $\tau_g$ is the temperature hyper-parameter.
This loss encourages all the group embeddings to be distinct.

\subsection{Prediction and Optimization}
For GI task, we conduct prediction on the ranking score ${y}_{u,g}$ of the user-group pair $(u,g)$, which is calculated by the inner product as:
\begin{equation}
{y}_{u,g} = \mathbf{e}_{u}\cdot \mathbf{e}_{g},
\end{equation}
where $\mathbf{e}_{u}$ is final layer user embedding as described in Eq.~(\ref{eq:user embed}) and $\mathbf{e}_{g}$ is the final layer group embedding as in Eq.~(\ref{eq3}).
Then we adopt the pairwise Bayesian Personalized Ranking (BPR) loss~\cite{bprloss} to optimize the prediction:
\begin{equation}
\mathcal{L}_{bpr}=\sum\limits_{(u,g,g')\in \mathcal{D}} -\log\sigma(\hat{y}_{u,g}-\hat{y}_{u,g'}),
\end{equation}
where $\mathcal{D}=\{(u,g,g')|g\in \mathcal{G}^{+}_{u}, g'\in \mathcal{G}\backslash\mathcal{G}^{+}_{u}\}$ is the training data and group set $\mathcal{G}^{+}_{u}$ contains all the groups joined by user $u$. 

Finally, we jointly optimize the recommendation task and the proposed cross-view self-supervised learning as follows:
\begin{equation}
\mathcal{L}= \mathcal{L}_{bpr} + \lambda(\mathcal{L}_{cssl}^{user} + \mathcal{L}_{reg}^{group}) + \lambda_\Theta\|\Theta\|^2_2,
\end{equation}
where $\lambda$ is the hyper-parameter to control the strength of SSL, and $\Theta$ is all trainable parameters in the framework, which is regularized by $\lambda_\Theta$. 
Adam~\cite{adam14} is chosen as the optimizer.

\subsection{Model Comparison}
In this section, we compare the proposed THC with convolution layers in the previous graph and hypergraph convolution based recommendation methods mathematically. 
We prove that both hypergraph convolution~\cite{mhcn} and GCN~\cite{GCN16} are special cases of THC.

\subsubsection{THC generalizes Hypergraph Convolution.}
By eliminating the non-linear activation and linear transformation for existing hypergraph convolution works~\cite{mhcn,hnn,dhcf},
we formulate a hypergraph convolution~\cite{mhcn} as: 
\begin{equation}\label{eq13}
\mathbf{E}^{(l+1)}=\mathbf{D}^{-1}\mathbf{H}\cdot\mathbf{B}^{-1}\mathbf{H}^\top\mathbf{E}^{(l)}.
\end{equation}
THC is equivalent to hypergraph convolution if we dismiss transition layer \textbf{Transition}() and set hyper-parameter $\gamma=0$ of Eq.~(\ref{thc}).

\subsubsection{THC generalizes LightGCN}
Since graph convolution is a special case of hypergraph convolution when each hyperedge connects only two nodes in hypergraph~\cite{BaiZT21}, THC as a general case of hypergraph convolution can also be regarded as a general case of graph convolution. Moreover, THC can be formulated as two consecutive graph convolution layers when each hyperedge connects multiple nodes.
We use the LGC layer in LightGCN~\cite{lightgcn20} to represent graph convolution, which is defined as follows:
\begin{equation}
    \mathbf{E}^{(l+1)} = (\mathbf{D}^{-\frac{1}{2}}\mathbf{A}\mathbf{D}^{-\frac{1}{2}})\mathbf{E}^{(l)},
\end{equation}
where $\mathbf{E}^{(l)}$ and $\mathbf{E}^{(l+1)}$ are the input and output of a graph convolution layer, and $\mathbf{A}$ is the adjacency matrix reflecting edges. To be more specific, two consecutive graph convolution layers for group-user-group information propagation are denoted as follows:
\begin{equation}
\begin{aligned}
    &\mathbf{E}_\mathbf{u}^{(l+1)} = (\mathbf{D}_{u}^{-\frac{1}{2}}\mathbf{A}^\top\mathbf{D}_{g}^{-\frac{1}{2}})\mathbf{E}_\mathbf{g}^{(l)},\\
    &\mathbf{E}_\mathbf{g}^{(l+2)} = (\mathbf{D}_{g}^{-\frac{1}{2}}\mathbf{A}\mathbf{D}_{u}^{-\frac{1}{2}})\mathbf{E}_\mathbf{u}^{(l+1)}.
\end{aligned}
\end{equation}
By combining the matrix forms of the two layers, we obtain updated group embeddings as follows:
\begin{equation}\label{eq16}
\begin{aligned}
    \mathbf{E}_\mathbf{g}^{(l+2)}& = (\mathbf{D}_{g}^{-\frac{1}{2}}\mathbf{A}\mathbf{D}_{u}^{-\frac{1}{2}})(\mathbf{D}_{u}^{-\frac{1}{2}}\mathbf{A}^\top\mathbf{D}_{g}^{-\frac{1}{2}})\mathbf{E}_\mathbf{g}^{(l)} \\
    &=\mathbf{D}_{g}^{-1}\mathbf{A}\mathbf{D}_{u}^{-1}\mathbf{A}^\top\mathbf{E}_\mathbf{g}^{(l)},
\end{aligned}
\end{equation}
where $\mathbf{A}\in\mathbb{R}^{|G|\times|U|}$ represents interactions between users and groups. Groups sharing a common user are connected together in the group hypergraph. We have its incidence matrix $\mathbf{H} = \mathbf{A}$, node degree matrix $\mathbf{D} = \mathbf{D}_{g}$ and hyperedge degree matrix $\mathbf{B} = \mathbf{D}_{u}$. Thus, Eq.~(\ref{eq16}) is equivalent to Eq.~(\ref{eq13}), which is a special case of THC.

\section{Experiment}

\begin{table}\caption{The statistics of datasets.}\label{tab:dataset}
\begin{tabular}{l|l|l|l}
\hline
\hline
\textbf{Dataset}           & \textbf{Steam} & \textbf{Beibei} & \textbf{Weeplaces} \\ \hline
\text{\# users}            & 19,608            & 11,487             & 8,550         \\ 
\text{\# groups}           & 46,587            & 4,035              & 8,535         \\ 
\text{\# items}            & 3,951             & 13,814             & 22,357          \\ 
\text{\# user-group edges} & 105,271           & 20,972             & 16,529         \\ 
\text{\# user-item edges}  & 1,209,979         & 105,210            & 152,258        \\ 
\text{Avg. \# groups/user} & 5.37              & 1.83               & 1.93         \\ 
\text{Avg. \# users/group} & 2.26              & 5.2                & 1.94         \\ 
\text{Avg. \# items/user}  & 61.71             & 9.16               & 18.44          \\ 
\text{Avg. \# users/item}  & 306.25            & 7.62               & 6.81          \\ \hline
\end{tabular}
\end{table}

\begin{table*}[!ht]\caption{Performance comparison on three datasets. 
}\label{tab:performance}
\begin{tabular}{l|cccc|cccc|cccc}
\hline
Dataset &
  \multicolumn{4}{c|}{Steam} &
  \multicolumn{4}{c|}{Beibei} &
  \multicolumn{4}{c}{Weeplaces} \\ \hline
Metric &
  R@10 &
  R@20 &
  N@10 &
  N@20 &
  R@10 &
  R@20 &
  N@10 &
  N@20 &
  R@10 &
  R@20 &
  N@10 &
  N@20 \\ \hline
LGCN &
  0.0938 &
  0.1236 &
  0.0734 &
  0.0803 &
  0.1042 &
  0.1329 &
  0.0741 &
  0.0821 &
  0.1176 &
  0.1506 &
  0.0705 &
  0.0790 \\
SGL &
  0.1290 &
  0.1565 &
  0.0879 &
  0.0856 &
  0.1084 &
  0.1377 &
  0.0779 &
  0.0883 &
  0.1192 &
  0.1486 &
  0.0707 &
  0.0784 \\
ENMF &
  0.1075 &
  0.1386 &
  0.0774 &
  0.0868 &
  0.1154 &
  0.1560 &
  0.0826 &
  0.0959 &
  0.1310 &
  0.1720 &
  0.0734 &
  0.0854 \\
HGNN &
  0.0398 &
  0.0515 &
  0.0407 &
  0.0463 &
  0.0945 &
  0.1279 &
  0.0864 &
  0.1070 &
  0.0967 &
  0.1367 &
  0.0468 &
  0.0605 \\
HCCF &
  0.0765 &
  0.0890 &
  0.0698 &
  0.0771 &
  0.0860 &
  0.1236 &
  0.0586 &
  0.0812 &
  0.0818 &
  0.1200 &
  0.0366 &
  0.0505 \\
DHCF &
  0.1747 &
  0.2082 &
  0.1549 &
  0.1724 &
  0.0828 &
  0.1261 &
  0.0629 &
  0.0876 &
  0.1080 &
  0.1466 &
  0.0633 &
  0.0780 \\
{LGCN+} &
  0.1019 &
  {\ul 0.2619} &
  0.0804 &
  0.1745 &
  0.0925 &
  0.1574 &
  0.0698 &
  0.1007 &
  {\ul 0.1668} &
  {\ul 0.2490} &
  {\ul 0.0824} &
  {\ul 0.1065} \\
GAT &
  0.1249 &
  0.1329 &
  0.1155 &
  0.1204 &
  0.0860 &
  0.1701 &
  0.1040 &
  0.1142 &
  \multicolumn{1}{r}{0.0350} &
  0.0716 &
  0.0175 &
  0.0298 \\
HGNN+ &
  {\ul 0.1752} &
  0.1753 &
  {\ul 0.1985} &
  {\ul 0.1986} &
  {\ul 0.1349} &
  {\ul 0.2003} &
  {\ul 0.1064} &
  {\ul 0.1410} &
  0.1500 &
  0.2337 &
  0.0723 &
  0.0956 \\\hline
GTGS &
  \textbf{0.2347} &
  \textbf{0.2903} &
  \textbf{0.2075} &
  \textbf{0.2335} &
  \textbf{0.2125} &
  \textbf{0.2970} &
  \textbf{0.1630} &
  \textbf{0.1931} &
  \textbf{0.2061} &
  \textbf{0.3090} &
  \textbf{0.0994} &
  \textbf{0.1292} \\ \hline
Improv. &
  33.96\% &
  10.84\% &
  4.53\% &
  17.57\% &
  57.52\% &
  48.28\% &
  53.20\% &
  36.95\% &
  23.56\% &
  24.10\% &
  20.63\% &
  21.31\% \\ \hline
\end{tabular}
\end{table*}

\subsection{Experimental Setup}
\subsubsection{Datasets} 
We conduct experiments on three real-world datasets: Steam~\cite{steam}, Beibei~\cite{beibei}, and Weeplaces~\cite{weeplaces}. Steam dataset includes users' transaction records and the discussion groups they joined on Steam online game store.
Beibei is the largest E-commerce platform for maternal and infant products in China. It records users' purchases and groups they join for group buying.
Weeplaces dataset includes users' group-traveling history and their check-in locations.
The statistics of three datasets are shown in Table \ref{tab:dataset}. 
For Steam and Beibei, we randomly select $70\%$ of all groups joined by each user for training and the remaining $30\%$ for testing. 
For Weeplaces, the split ratio is $80\%$ for training and $20\%$ for testing. We further split $20\%$ validation set from the training set for hyper-parameter tuning.
Our source code and the three datasets are released online \footnote{https://github.com/mdyfrank/GTGS}.

\subsubsection{Baselines}
We compare GTGS with the following baselines:
\begin{itemize}[leftmargin=*]
    \item \textbf{LGCN}~\cite{lightgcn20}. This is the state-of-the-art recommendation method based on GCN~\cite{GCN16} by removing feature transformation and nonlinear activation.
    \item \textbf{SGL}~\cite{sgl21}. This recommendation framework performs contrastive learning on LGCN to augment node representations by leveraging additional self-supervised signals during training.
    \item \textbf{ENMF}~\cite{enmf20}. It is based on a neural matrix factorization architecture leveraging mathematical optimization to train the model efficiently without sampling data for recommendation.
    \item \textbf{HGNN}~\cite{hnn}. It is a method applying hypergraph convolution on a graph for node representation learning. We calculate inner production between user nodes and group nodes as the prediction scores with BPR loss as the objective for recommendation. 
    \item \textbf{HCCF}~\cite{hccf}. In this hypergraph learning based recommendation method, contrastive learning is applied to jointly capture local and global collaborative relations.
    \item \textbf{DHCF}~\cite{dhcf}. It introduces hypergraph convolution into dual-channel learning for recommendation.
    \item \textbf{LGCN+}. This is a GCN-based recommendation model adapted for GI task. It consists of two LGCN parts to conduct item-user and user-group information propagation consecutively.
    \item \textbf{HGNN+}. This recommendation model is based on HGNN. We adapt it for GI task by employing hypergraph convolution on user-item and user-group bipartite graphs successively.
    \item \textbf{GAT}~\cite{gat18}. This work leverages a multi-head self-attention architecture to capture influences from each node in graph structures. To adapt it for GI task, we deploy two consecutive graph attention networks for item-user and user-group information aggregation and use pair-wise BPR loss as the objective function. 
\end{itemize}
Since those bipartite recommendation baselines are not designed for graphs with multiple types of interactions, we deploy them with only user-group interactions such that they are adapted to GI task.

\subsubsection{Model Settings}
The number of THC layers $L$ is set to $1$ in our experiments.
Learning rate is set to $0.05$ for Steam/Beibei and $0.005$ for Weeplaces.
We set regularization strength $\lambda_{\theta}$ to 1e-7 for Beibei/Weeplaces and 1e-5 for Steam.
Embedding size $d$ is set to 64 for all the datasets. 
We adopt full-batch training. Early stopping is utilized in all experiments to cope with the over-fitting problem.
\subsubsection{Evaluation Metrics}
We evaluate GI task by ranking the test groups with all non-interacted groups of users. 
And we adopt Recall$@ \{10,20\}$ and NDCG$@\{10,20\}$ as evaluation metrics. 

\subsection{Overall Performance Comparison}
Overall comparison results are shown in Table~\ref{tab:performance}. 
The best results are in boldface, and the second-best results are underlined. The improvement is calculated by subtracting the best performance value of the baselines from that of GTGS and then using the difference to divide the former.
We summarize the following key observations:
\begin{itemize}[leftmargin=*]
    \item The proposed GTGS method achieves the best results and outperforms all the baseline methods by up to $57.52\%$ in the three datasets. We hypothesize these large stable gains result from the abundant and balanced user-item and user-group interactions.
    \item The baselines integrating both user-item and user-group interaction information, such as LGCN+ and HGNN+, are better than most bipartite recommendation models. It indicates that users' interest in items is able to benefit group identification as side information rather than harmful noise in most cases.
    However, LGCN+ and GAT are still unable to capture relationships among multiple users or groups synchronously through hypergraphs, thus being worse than GTGS. 
    This observation justifies the necessity of hypergraph learning for GI task.
    \item 
    Although HGNN+ utilizes hypergraphs to learn user-group and user-item relations, it models these two types of relations independently and neglects the intrinsic consistency between users' preferences for items and groups.
    Compared with it, GTGS is specifically designed for GI task with transitional hypergraph convolution and cross-view self-supervised learning, which is a better and more stable framework.
\end{itemize}

\subsection{Ablation Study}
\subsubsection{Hypergraph construction}
\begin{figure}[h]
    
    \begin{subfigure}{0.155\textwidth}
    \includegraphics[width=\textwidth]{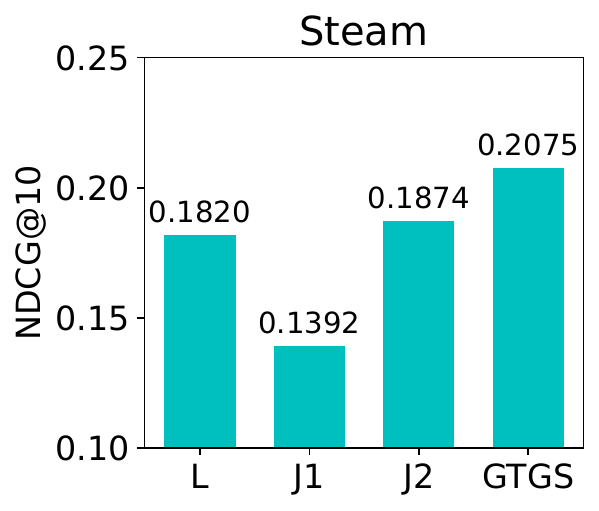}
    \end{subfigure}
    \hfill
    \begin{subfigure}{0.155\textwidth}
    \includegraphics[width=\textwidth]{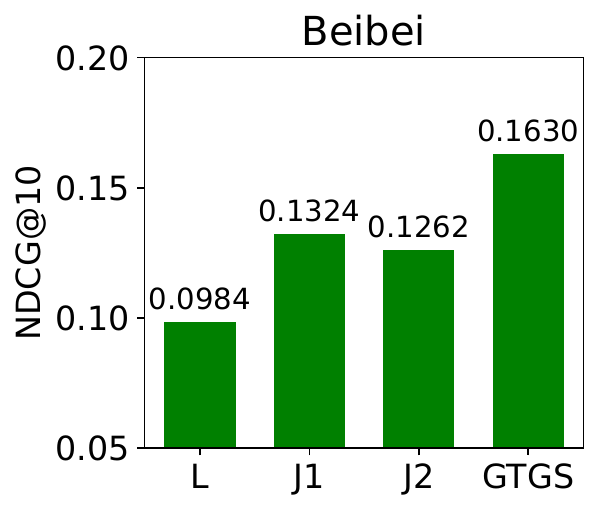}
    \end{subfigure}
    \hfill
    \begin{subfigure}{0.155\textwidth}
    \includegraphics[width=\textwidth]{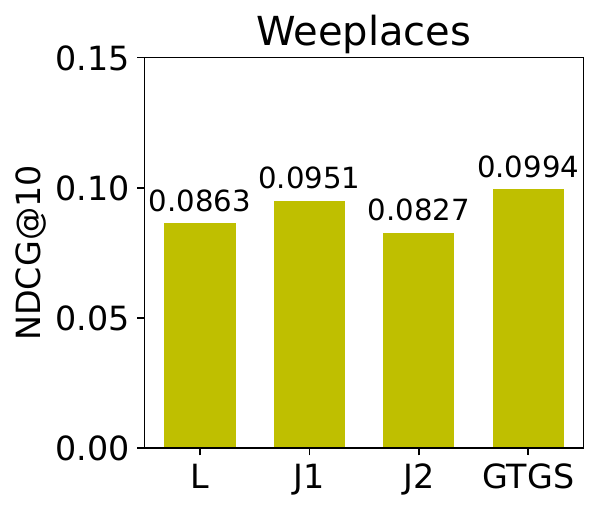}
    \end{subfigure}
        \caption{Performance of GTGS w.r.t. different hypergraph constructions.}
  \label{fig:HyperConstruct}
\end{figure}

We demonstrate the performance of GTGS with different item-view user hypergraph constructions on the three datasets in Fig.~\ref{fig:HyperConstruct}.
GTGS utilizes item nodes as hyperedge to construct the item-view user hypergraph.
In addition, we investigate three other variants of GTGS: (L) Instead of constructing the item-view user hypergraph, item information is aggregated to user nodes through user-item interactions by one GCN layer; 
(J1) both item nodes and group nodes are used as hyperedges among user nodes, and hypergraph convolution is employed on both item hyperedges and group hyperedges simultaneously;
(J2) same as J1, but hypergraph convolution is first employed on item hyperedges and then on group hyperedges.
We only show the result on NDCG since the pattern on Recall is the same. 

We observe that GTGS performs the best 
on three datasets. This justifies the superiority of GTGS with simply item hyperedges that perform best among all item-view user hypergraph constructions.
The poor performance of L verifies the limitation of bipartite edges in capturing real-world non-bipartite relations.
Since user-item interactions are much denser than user-group interactions in Steam dataset, the two-stage hypergraph convolution in J2 is able to distinguish the information propagated by item and group hyperedges, which results in better performance than J1. However, given user nodes have been used as hyperedges in the group hypergraph, adding group hyperedges to the item-view user hypergraph increases the risk of over-smoothing between users and groups, so both J1 and J2 are worse than GTGS.

\subsubsection{Transitional Hypergraph Convolution Settings}
\begin{figure}[h]
    \begin{subfigure}{0.155\textwidth}
    \includegraphics[width=\textwidth]{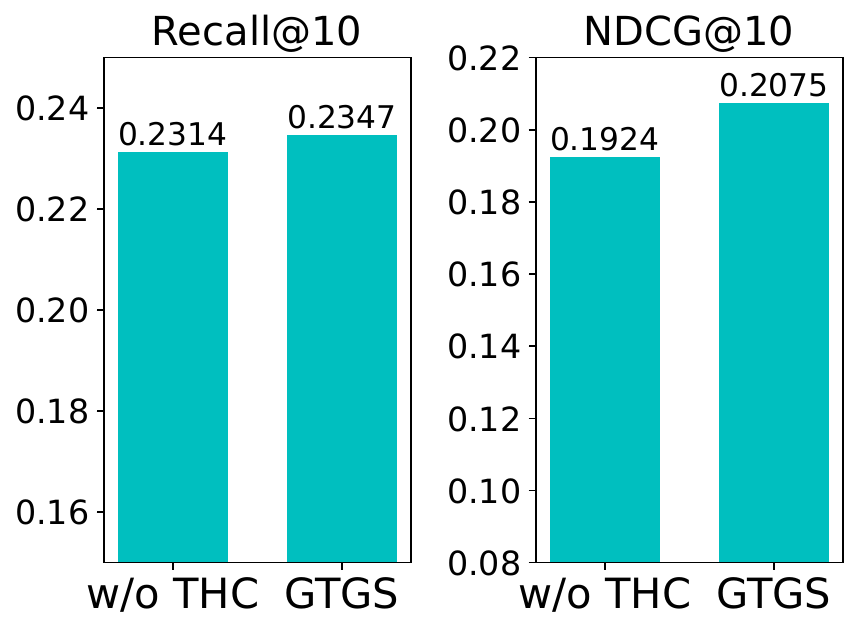}
    \caption{Steam}
    \end{subfigure}
    \hfill
    \begin{subfigure}{0.155\textwidth}
    \includegraphics[width=\textwidth]{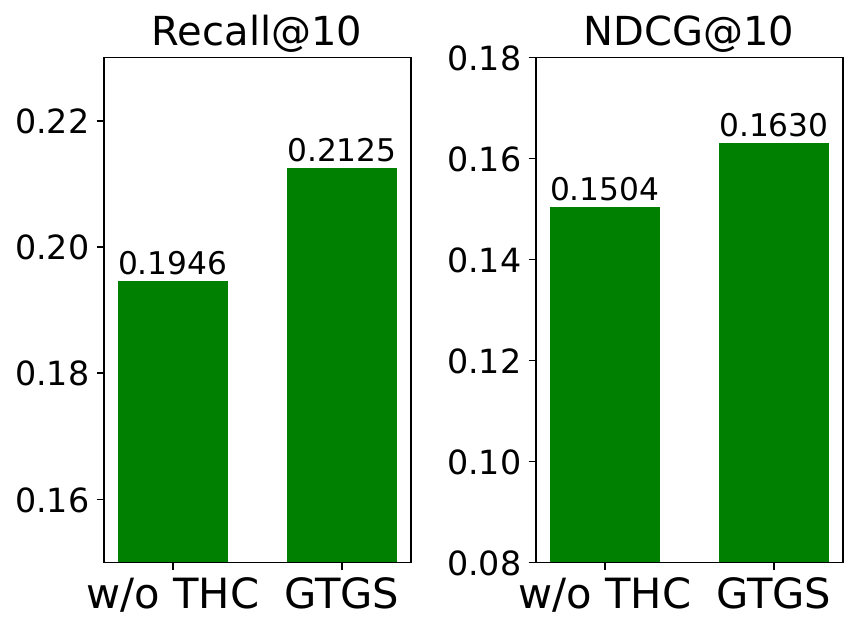}
    \caption{Beibei}
    \end{subfigure}
    \hfill
    \begin{subfigure}{0.155\textwidth}
    \includegraphics[width=\textwidth]{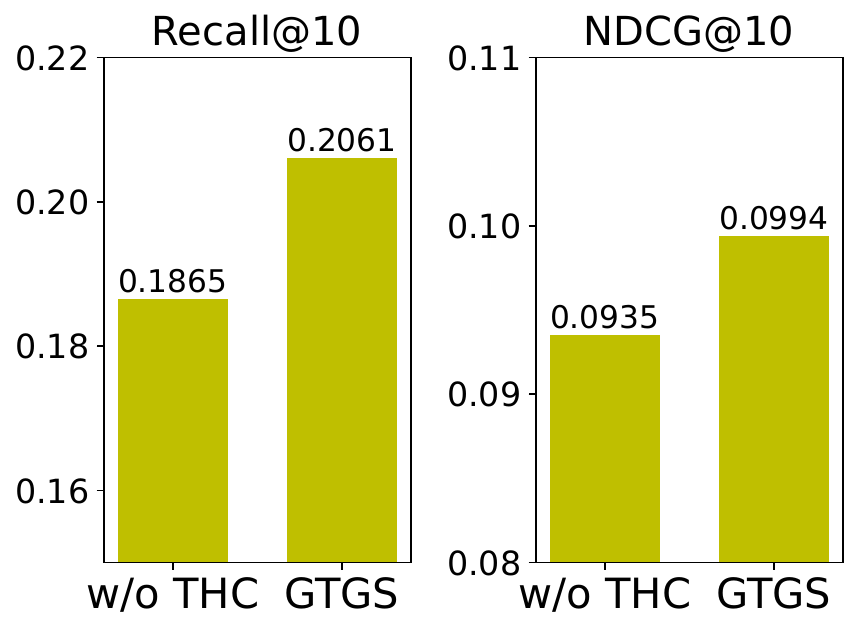}
    \caption{Weeplaces}
    \end{subfigure}
        \caption{Performance of GTGS with and without THC.}
  \label{fig:THCablation}
\end{figure}

To verify the effectiveness of THC, we show the performance of GTGS with and without THC in Fig.~\ref{fig:THCablation}. For the model without THC, we set transition intensity $\gamma=0$ for all THC layers.
We can observe that the performance is consistently improved by THC on all three datasets. 
And the improvement is more obvious in the sparser datasets, such as Beibei and Weeplaces.
In these datasets, members' interests in items are more distinct in each group, thus helping the model more accurately capture the characteristics of their group.   
Therefore, we believe that including users' item preferences in group representations by THC is a better way for GI task.

\subsubsection{Self-supervised learning settings}
\begin{table}[h]\caption{Ablation study on self-supervised learning settings.}\label{tab:ssl}
\resizebox{0.48\textwidth}{!}{%
\begin{tabular}{l|cccccc}
\hline
Dataset & \multicolumn{2}{c}{Steam}      & \multicolumn{2}{c}{Beibei}     & \multicolumn{2}{c}{Weeplaces}         \\ \hline
Metric    & R@10   & N@10   & R@10   & N@10   & R@10   & N@10   \\ \hline
w/o SSL  & 0.2214 & 0.1829 & 0.1841 & 0.1470 & 0.1821 & 0.0931 \\
w/o user  & 0.2242 & 0.2037 & 0.1865 & 0.1367 & 0.1831 & 0.0939 \\
w/o group & 0.2325 & 0.1888 & 0.1900 & 0.1482 & 0.1964 & 0.0989 \\
\hline
GTGS    & \textbf{0.2347} & \textbf{0.2075} & \textbf{0.2028} & \textbf{0.1526} & \textbf{0.2061} & \textbf{0.0994} \\ \hline
\end{tabular}%
}
\end{table}
Another ablation study is further made to investigate the effectiveness of self-supervised learning.
We design three different variants of GTGS. 1) \textit{w/o SSL} is without both CSSL and group-based regularization during the training of GTGS, so the model is optimized only by the BPR loss.
2) \textit{w/o user} is GTGS without the CSSL to maintain the consistency between item-view and group-view user representations. 
And 3) \textit{w/o group} is GTGS without the group-based regularization to enhance the distinction of group representations. 
The results are demonstrated in Table~\ref{tab:ssl}. We observe that:
\begin{itemize}[leftmargin=*]
    \item The best setting in general is to apply CSSL and group-based regularization to optimize both user and group representations. This verifies the effectiveness of CSSL applied in GTGS.
    \item Compared with those variants without the CSSL (i.e., \textit{w/o SSL} and \textit{w/o user} in Table~\ref{tab:ssl}), adding it to the model consistently improves the performances in all datasets, which justifies the importance of maintaining the consistency between item and group preferences during the user representation learning.
    \item Removing the group-based regularization will drop the performance in NDCG@10 largely in the Steam dataset, which has more groups and denser user-group interactions. This indicates that distinction among embeddings is crucial for learning representations of densely connected nodes by hypergraph convolution. Since the embeddings of connected nodes tend to be close in the vector space after convolution~\cite{insightgcn18}, it is more important to prevent the features of densely connected nodes from being excessively similar and indistinguishable in such a dataset.
\end{itemize}

\subsection{Cross-view SSL Analysis}
\begin{figure}[h]
    \begin{subfigure}{0.16\textwidth}
    \includegraphics[width=\textwidth]{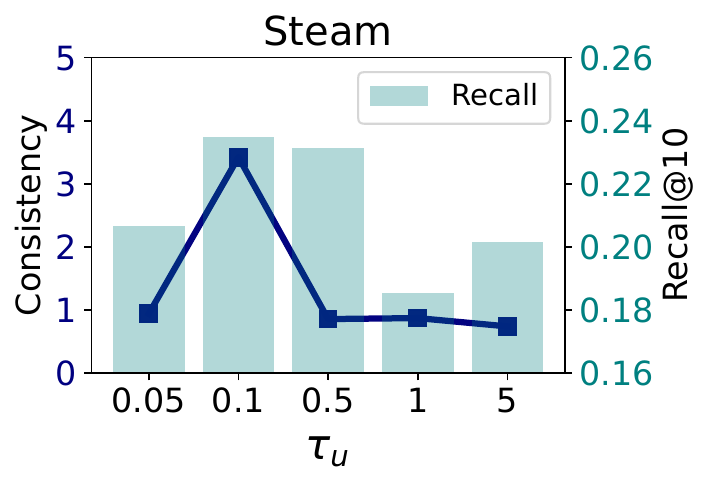}
    \end{subfigure}
    \hspace{-15mm}
    \hfill
    \begin{subfigure}{0.16\textwidth}
    \includegraphics[width=\textwidth]{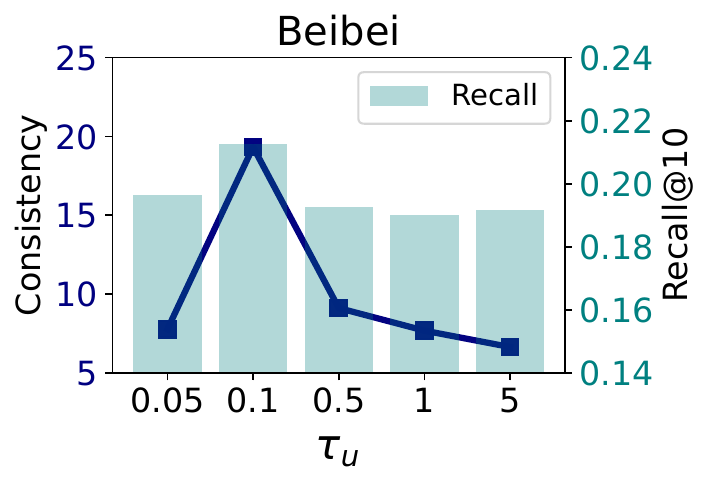}
    \end{subfigure}
    \hspace{-15mm}
    \hfill
    \begin{subfigure}{0.16\textwidth}
    \includegraphics[width=\textwidth]{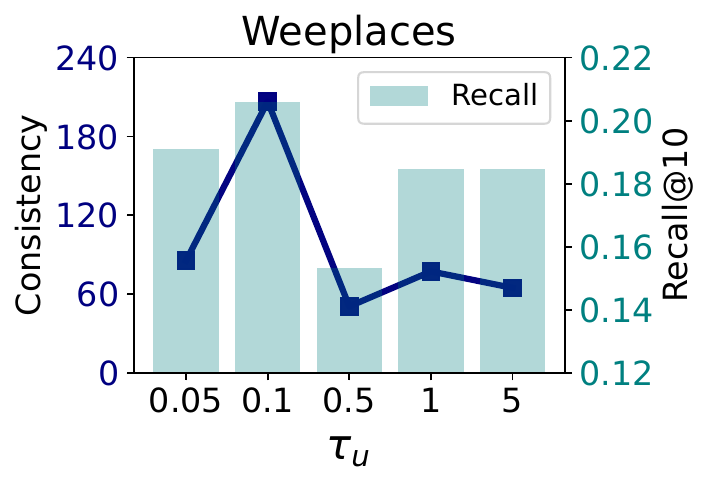}
    \end{subfigure}
    
  \caption{Average consistency of user embeddings w.r.t CSSL temperature $\tau_\mathbf{u}$. Histograms denote performances in Recall.}
  \label{fig:SSLstudy_user}
\end{figure}
    
As aforementioned, CSSL encourages the intrinsic consistency between item-view and group-view user embeddings for learning comprehensive user representation.
We further analyze the influences of user embedding consistency on performance. We define the consistency between item-view and group-view user embeddings as follows:
\begin{equation}
    \text{Consistency} = \frac{\mathop{\mathbb{E}}_{u\sim p_{user}}\text{sim}(\mathbf{E}_\mathbf{u}^{i,(L)},\mathbf{E}_\mathbf{u}^{g,(L)})}{\mathop{\mathbb{E}}_{u,v\sim p_{user}}\text{sim}(\mathbf{E}_\mathbf{u}^{i,(L)},\mathbf{e}_\mathbf{v}^{i,(L)})},
\end{equation}
where $p_{user}(\cdot)$ is the distribution of user data and $\text{sim}(\cdot)$ denotes the consine similarity. This consistency score is the ratio of the expected similarity between the two views of a user to the expected similarity between any two users in the dataset. 
A larger consistency means that the two views of each user learned by GTGS are more consistent.
Consistency among user representations learned by CSSL with different temperatures $\tau_{u}$ is shown in Fig.~\ref{fig:SSLstudy_user} together with corresponding model performances.

We observe that the model has the best performance when consistency achieves the peak, which indicates that the proposed CSSL is able to boost recommendation performance by enhancing the agreement between item-view and group-view user embeddings. Besides, the relatedness between consistency and performance is more obvious in smaller Beibei and Weeplaces datasets. Since each user only interacted with several items in such datasets, her group preference can be clearly revealed by her interests in items. This justifies the CSSL in our design.
\subsection{Embedding Analysis}
\begin{figure}[h]
    \begin{subfigure}{0.16\textwidth}
    \includegraphics[width=\textwidth]{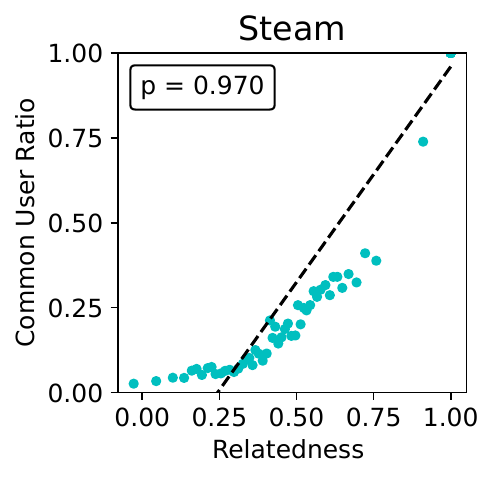}
    \end{subfigure}
    \hspace{-15mm}
    \hfill
    \begin{subfigure}{0.16\textwidth}
    \includegraphics[width=\textwidth]{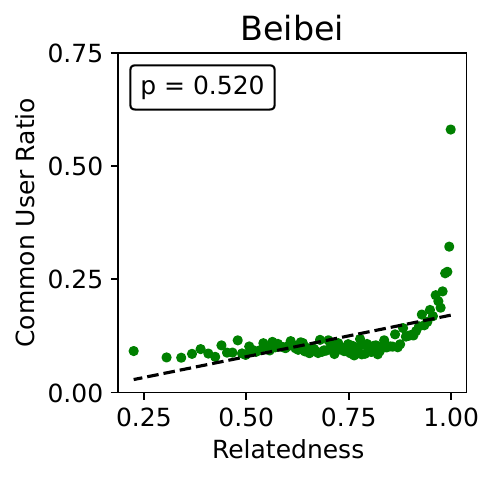}
    \end{subfigure}
    \hspace{-15mm}
    \hfill
    \begin{subfigure}{0.16\textwidth}
    \includegraphics[width=\textwidth]{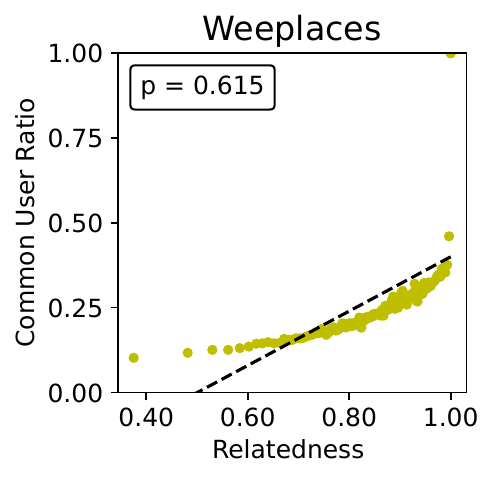}
    \end{subfigure}
    
  \caption{Common user ratio w.r.t relatedness of group pairs based on group embeddings.}
  \label{fig:EmbeddingAnalysis}
\end{figure}

We conduct analyses of the group embeddings learned by GTGS to verify that those embeddings can reflect the user composition of groups through hypergraph learning.
We retrieve all pairs of groups. We calculate their relatedness and corresponding common user ratio. The ratio for each group pair $(g_a,g_b)$ is computed as:
\begin{equation}\label{eq:common user ratio}
r_{ab} = \frac{|\mathcal{N}^{(u)}_{g_a}\cap\mathcal{N}^{(u)}_{g_b}|}{|\mathcal{N}_{g_a}^{(u)}\cup\mathcal{N}^{(u)}_{g_b}|},
\end{equation}
where $\mathcal{N}_{g_a}^{(u)}$ and $\mathcal{N}_{g_b}^{(u)}$ is the set of users in group $g_a$ and $g_b$, respectively. For a simple illustration purpose, we sort all group pairs w.r.t. the pair-wise cosine similarity as the relatedness score and split them into $100$ equal size subsets, and represent each subset as the average scores.
The scatter plots between relatedness and common user ratio sharing ratio on three datasets are shown in Figure~\ref{fig:EmbeddingAnalysis}. 
We also draw a regression line and compute the Pearson correlation coefficient $p$ for each dataset.  
We observe an obvious overall tendency that the common user ratio increases with the growth of relatedness. This tendency indicates that two groups sharing more common members have higher relatedness based on their embeddings, which justifies the efficacy of hypergraph learning in modeling group representations.

\subsection{Cold-start Group Identification}
\begin{figure}[h]
    \vspace{-8.5mm}
    \begin{subfigure}{0.5\textwidth}
    \centering
    \includegraphics[width=\textwidth]{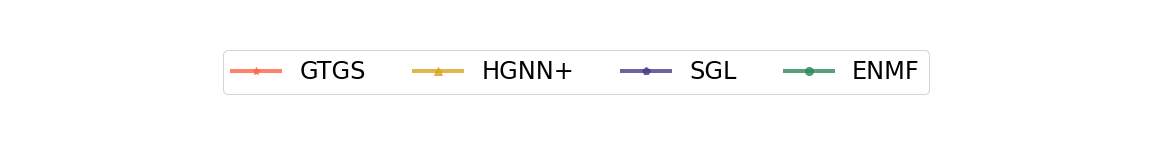}
    \end{subfigure}
    \vspace{-8mm}
    
    \begin{subfigure}{0.235\textwidth}
    \includegraphics[width=\textwidth]{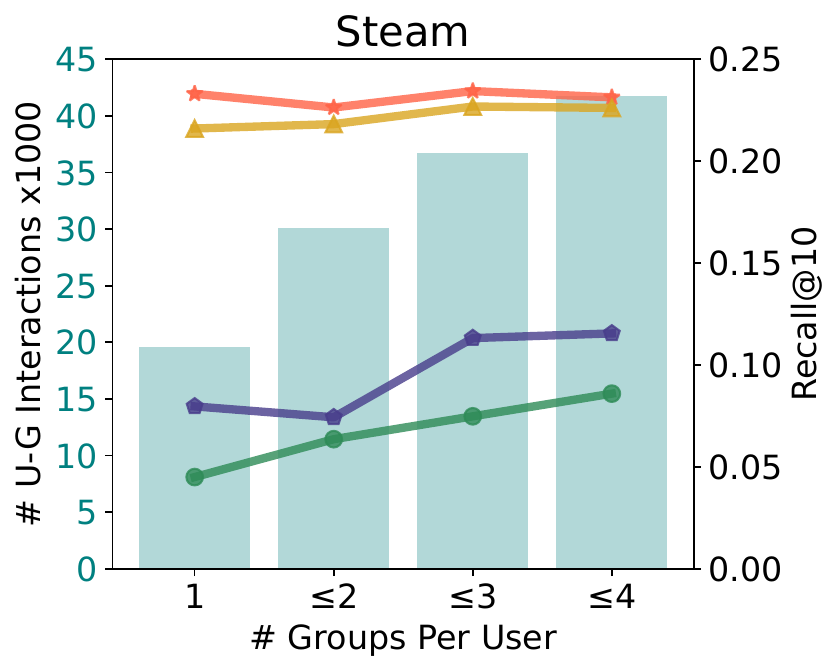}
    \label{fig:steam_cold}
    \end{subfigure}
    \hspace{-1mm}
   \begin{subfigure}{0.235\textwidth}
    \includegraphics[width=\textwidth]{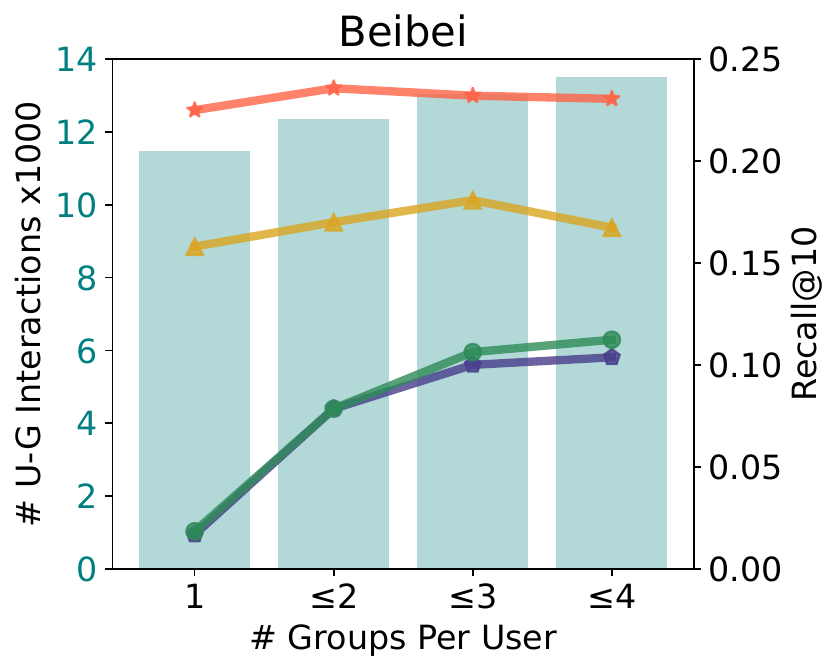}
    \label{fig:beibei_cold}
    \end{subfigure}
    \caption{Cold-start performance of different methods. 
    The background histograms denote the number of user-group interactions left in the training set with different thresholds. 
The solid lines indicate the performances of different methods with respect to different thresholds.
}
    \label{fig:cold}
\end{figure}
As shown in Table~\ref{tab:dataset},  many users have few group interactions, and the cold-start issue in GI task is severe. Hence, we conduct a detailed analysis regarding the ability of GTGS to tackle cold-start group recommendation. We randomly remove some user-group edges for each user in the training set such that the number of neighbor groups of each user is no greater than a threshold $k$. 
For comparison, we choose SGL, ENMF, and HGNN+ as the three baseline models, and perform the experiments with threshold $k\in\{1,2,3,4\}$. The threshold indicates the maximum number of groups per user.
The results are shown in Figure~\ref{fig:cold}.
We report the Recall performance with respect to different thresholds as the solid line and the number of user-group interactions in the background
histograms. First, We observe that GTGS as the method leveraging user's item preferences for GI task significantly outperforms SGL and ENMF as the bipartite user-group recommendation models. The reason is that GTGS can leverage both the item and group interactions to learn node embeddings. Despite the few group participation of users, item interactions complement the cold-start issue. Even if only one user-group edge is left, GTGS is able to achieve outstanding performance in Steam and Beibei datasets. Secondly, although HGNN+ also utilizes hypergraph learning to model user-group relations with user-item relations, it ignores the consistency between users' preferences for items and groups. Thus it relies on user-group edges more heavily and still performs worse than GTGS in Beibei dataset with sparse user-item interactions. 

Additionally, we observe that sometimes GTGS even performs better with fewer user-group edges, such as $k=1$ in Steam dataset.
We hypothesize that when the number of groups per user is few, GTGS can well characterize the group interests of users from their item interactions. 
Therefore, it verifies the GTGS is a better framework to comprehensively integrate item and group information for users and can successfully complete the GI task.

\section{Conclusion}
In this paper, we propose a novel framework GTGS for GI task. GTGS leverages hyperedges to express the union semantics in users and groups for recommendation. A THC layer is devised to transmit information on members' item preferences to groups, such that users' interests in items serve as prior knowledge during group identification. In addition,
CSSL is leveraged to guarantee the intrinsic consistency between
item and group preferences for each user, and group-based regularization is used to enhance the distinction among group representations.
We conduct extensive experiments and detailed
analyses on three datasets to verify the effectiveness of GTGS. In the future, we may explore how to design other variants of transition layers in THC for the transition of prior knowledge, so that we can improve the functionality of THC layers.

\section{Acknowledgements}
This work is supported by National Key R\&D Program of China through grant 2022YFB3104700, NSFC through grants U21B2027, 62002007, 61972186 and 62266028, NSF under grants III-1763325, III1909323, III-2106758, and SaTC-1930941, S\&T Program of Hebei through grants 20310101D and 21340301D, Beijing Natural Science Foundation through grant 4222030, Yunnan Provincial Major Science and Technology Special Plan Projects through grants 202202AD080003, 202103AA080015 and 202203AP140100, grant 202301AS070047 under General Projects of Basic Research in Yunnan Province, and the Fundamental Research Funds for the Central Universities.

\bibliographystyle{ACM-Reference-Format}
\bibliography{sample-base}

\end{document}